\documentclass[aps,pre,epsfig,floats,twocolumn,superscriptaddress,amssymb,amsmath,floatfix,showpacs]{revtex4}
\usepackage{graphicx}
\usepackage{bm}  
\usepackage{amssymb}
\usepackage{color}
\usepackage{amscd}

\begin{document}

\title{ Hydrodynamical random walker with chemotactic memory}

\author{H. Mohammady, B. Esckandariun and A. Najafi}
\email{najafi@znu.ac.ir} 
\affiliation{Physics Department, University of Zanjan, Zanjan 45371-38791, Iran}

\date{\today}

\begin{abstract}
A three-dimensional hydrodynamical model for a micro random walker is combined with the idea of 
chemotactic signaling network of E. coli. 
Diffusion exponents, orientational correlation functions  
and their dependence on the geometrical and dynamical parameters of the system are analyzed numerically. 
Because of the chemotactic memory, the walker shows  superdiffusing displacements  in all directions 
with the largest diffusion exponent for a direction along the food gradient.        
Mean square displacements and orientational correlation functions show that the chemotactic memory washes out all the signatures due to the geometrical asymmetry  of the walker and statistical properties are asymmetric only with respect to the 
direction of food gradient. For different values of the memory time, the Chemotactic index (CI) is also 
calculated.
\end{abstract}


\date{\today}

\maketitle 
\section{Introduction}
Random walk as a general mathematical tool can describe a large class of biophysical systems such as 
the motion of Brownian colloidal particles and the motion of biological 
self-propelled microorganisms \cite{RWrev,bergbook,reinforcedRW,statchemo,biasedswiming,vafabakhsh}. 
Passive colloidal particles in response to the thermal forces randomly change their directions but  microorganisms sense the directions and overcome the thermal randomness to reach a predefined target point. 
Microorganisms do not have a complex intelligence in the form of a brain that can process the  
complicated signals from different senses. 
Chemotaxis is a mechanisms that microscopic organisms use to detect their right tracks and navigate 
toward their targets \cite{berg,adler,naturereview,kaup}. 
The phenomenon of chemotaxis has inspired extensive research both due to 
its direct  biological relevance \cite{bray,BergEC,frankchem} and 
also because of the practical needs for designing artificial nanorobots that can sense the direction \cite{mit,molmot}.    
As the swimming of bacteria takes place in aqueous media, the inertialess condition in microscopic 
world results a  peculiar fluid dynamic problem\cite{Purcell}. 
From the other hand, at the scale of micrometer, the fluctuations are the non negligible part of the physics and any kind 
of modeling should take into account the effects due to randomness.  
Mathematical description of chemotaxis in terms of random walk requires a knowledge about the values 
of jumping displacements and rotations and  their transition probabilities. 

So far, theoretical studies on random walk modeling for microorganisms do not 
consider the hydrodynamical details and also the mechanism of chemotactic memory  in a unified model.  
In this article, we aim to construct a model that takes into account  the details   
of hydrodynamical displacements,  mechanism of chemotaxis memory  and also the physics of fluctuations.   
This model will exhibit  detailed features of the motion that provides   a  coupling between the geometrical parameters of the walker with 
the conformational rates. Geometrical parameters contribute through the hydrodynamic part and the 
dynamical parameters will enter through the chemotactic memory mechanism. 

The structure of this article is as follows: In section II we introduce the hydrodynamical details of the system and the chemical mechanism for the 
memory of the swimmer will be studied in section III. Finally the statistical results based on numerical investigations will be presented in section IV.
\begin{figure}
\centering
\includegraphics[width=0.95\columnwidth]{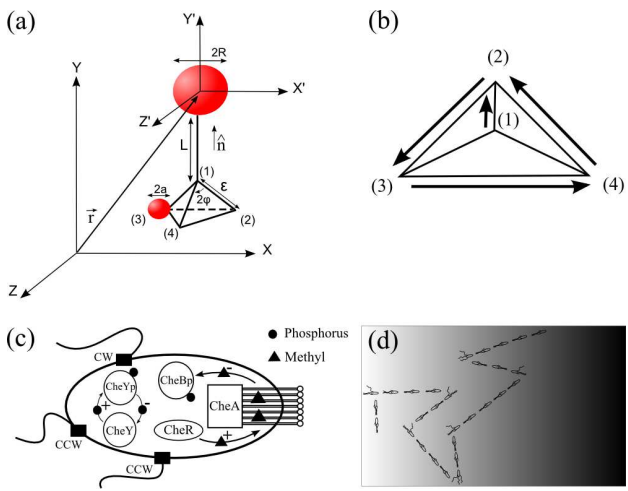}
\caption{ 
(a) A sphere with radius $R$ models the body of a bacterium and a moving small sphere in 
resemblance with flagellum, provides the driving force. The hydrodynamical calculations are 
done in a co-moving frame of reference. 
(b) A set of jumps are chosen as CW rotations. The rate of this CW rotations are determined by the chemotactic response function while the rate for other jumps are given randomly.
(c) A simplified picture of the inter cellular chemical network in E. coli. 
Two important processes of phosphorylation and methylation take place inside cell. Phosphorylated 
CheY-P enzymes produced by cheA (enzymes connected to receptors), 
are responsible to enhance the rotational direction of flagella and subsequently force bacterium 
to tumble. Methylation level  
of receptors from one hand and the concentration of the food from other hand, change the activity of receptors and enhance the phosphorylations process. (d)  Subsequent runs and tumbles would 
lead the bacterium to find the source of food.  
}
\label{fig1}
\end{figure}

\section{Hydrodynamical Model}
Our goal in this article is to combine the idea of chemical memory with a hydrodynamical model of a walker. 
Now let us introduce the hydrodynamical details of our system.  
Inspired by a Bacterium, Fig.~\ref{fig1}(a) shows the body of our walker that  is 
modeled by a sphere of radius $R$. The driving force is modeled by a mobile small sphere with 
radius $a$ ($a\ll R$). These two spheres are connected by an arm with negligible diameter. 
This model resembles the geometry of a bacterium with a single tail. 
As shown in this figure, and in a reference frame connected to the large sphere, jumps of 
this small sphere between $4$ vertices of a pyramid will construct all 
internal discontinues jumps of the 
walker. The apex of this pyramid is a point  with a distance $L$ apart from
the large sphere and is chosen as state $(1)$. The other $3$ states are located on the base of this 
pyramid. The apex angle is $2\phi$ and the apex sides are $\varepsilon$.    
The base of this pyramid is an equilateral triangle with sides $2\varepsilon\sin\phi$. The 
angle $\varphi$ may resemble the amplitude of flagellum undulations.
The hydrodynamic question that we need to address here, is the   
differential change of the position and orientation of the system for an internal jump.
For a fluid with viscosity $\eta$ and at the condition of micron scale, the inertialess stokes equation $\eta\nabla^2{\bf u}-\nabla P=0$, written for the 
fluid velocity ${\bf u}$ and pressure $P$, describes the dynamics. The condition of 
incomprehensibility, should also be considered.   
A prescribed motion corresponding to a jump of 
the  small sphere, will enter to the dynamics through  the boundary conditions. 
Solving the Stokes equation with the corresponding boundary conditions  would result the 
dynamical properties of the large sphere during an internal jump. 
Calculations similar to the details presented elsewhere, reveals the dynamical results \cite{molhun}.
To summarize the hydrodynamical results, let's denote the relative 
speed  of small sphere with respect to the large sphere in each jump  by $v$. Now the 
differential displacement and rotation of the large sphere in the laboratory frame 
for a jump from state $(i)$ to state $(j)$  read:
$$
\Delta\vec{r}^{H}_{ij}={\cal R}( \hat{n})\delta\vec{x}_{ij},~~~\Delta\hat{n}^{H}_{ij}={\cal R}( \hat{n})\delta\vec{\omega}_{ij}\times\hat{n},
$$
where $\hat{n}$ represents the  directer vector of the walker and 
${\cal R}(\hat{n})$ is an appropriate rotation matrix that transform the comoving frame of reference to the frame of laboratory. The comoving frame is shown in 
Fig.~\ref{fig1}(a). The  differential rotations and 
 displacements in the comoving frame are given by:
\begin{equation}
\vec{\omega}_{12}=\left (
 \begin{matrix}
\frac{-2\alpha}{\sqrt{3}}\sin\varphi\nonumber\\
0\nonumber\\
-2\alpha\sin\varphi\nonumber\\

\end{matrix}\right),~~~
\vec{\omega}_{24}=\left (
 \begin{matrix}
3\alpha\nonumber\\
0\nonumber\\
\alpha\nonumber\\
\end{matrix}\right).~~~
\end{equation}
\begin{equation}
\delta\vec{x}_{12}=\left (
 \begin{matrix}
-\delta\sin\varphi\nonumber\\

\delta '\nonumber\\
\frac{\delta}{\sqrt{3}}\sin\varphi\nonumber\\
\end{matrix}\right),~~~
\delta\vec{x}_{24}=\left (
 \begin{matrix}
\frac{\delta}{2}\nonumber\\
0\nonumber\\
\frac{-\sqrt{3}\delta}{2}\nonumber\\
\end{matrix}\right),~~~
\end{equation}
where the parameters are defined as:
\begin{align}
\alpha&=\frac{3}{8}~(\frac{\varepsilon}{R})~(\frac{a}{R})(1+\frac{L}{R}),~~~\delta=(\frac{\varepsilon}{R}) ~(\frac{a}{R})~(1-\frac{3}{4}\frac{R}{L}),\nonumber\\
\delta '&=(\frac{\varepsilon}{R}) ~(\frac{a}{R})~(1-\frac{3}{2}\frac{R}{L})~\sqrt{1-\frac{4}{3}\sin ^{2}\varphi}.\nonumber
\end{align}
Please note that the results are presented for a walker with $\varepsilon\ll L$. 
The differential changes are given only for two jumps, a jump  started from the apex and a 
jump in the base face of the pyramid.  
Other jumps can be obtained from these two special jumps by applying 
the appropriate rotation matrices. Symmetry requires that $\vec{\omega}_{ij}=-\vec{\omega}_{ji}$ and $\delta\vec{x}_{ij}=-\delta\vec{x}_{ji} $. 
A hydrodynamic time scale can be defined as: $\tau^H=\varepsilon/v$. This is 
the time for jumps that start farm the apex of pyramid, the time for other jumps are given by: 
$2\tau^H\sin\varphi$. 

In the next section we first introduce the chemotactic memory of a bacterium then show how can we combine the idea of chemotactic memory with the above introduced hydrodynamical swimmer. 

\section{Chemotactic memory}
To have a plan for internal jumps, we use the chemotactic strategy that bacteria use to navigate. 
Among different microorganisms, the chemical network responsible for chemotactic signaling is well
understood and studied in E. coli \cite{ecoli1,ecoli2}. 
Running state of this bacterium is due to the CCW (counter clockwise) rotation of flagella and 
changing the flagellar rotational state to CW (clockwise) will result a tumble.   
Chemical signals inside cell, controls the frequency of these running and tumbling states.   
A very simplified picture of different proteins involved in the 
chemical signal transduction pathway is shown in Fig.~\ref{fig1}(c). 
Two important processes of phosphorylation and methylation take place inside cell. Phosphorylated 
CheY-P enzymes produced by cheA (enzymes connected to receptors), 
are responsible to enhance the above mentioned frequency. Methylation level  
of receptors from one hand and the concentration of the food from other hand, change the activity of receptors and enhance the phosphorylation process \cite{vladimirov1,vladimirov2,vladimirov3}.
The methylation process 
provides a kind of chemical memory for the cell and allows the organism 
to compare the current local value of food 
concentration by its value at past. Depending on the parameters, A bacterium with this sort of memory will have chance to find a way to reach a point with maximum value of food concentration, see 
Fig.~\ref{fig1}(d).

Now we want to combine the idea of chemotactic memory with the details of 
hydrodynamic jumps of the walker. Our modeling is based on stochastic jumps. 
Let denote the transition probability for jump from 
state $(i)$ to state $(j)$ by $P_{ij}$. In the case with 
$P_{ij}=1/3$, we will have a random walker that has no chance to sense the direction of gradient. 
What we 
want to consider, is a sort of an intelligent walker that can dynamically change 
its jumping probabilities. 
In comparison with E. coli, we first define a set of jumps that corresponds to a CW rotation. 
We define all the following jumps as CW jumps\cite{cwrotation}:
$$
CW ~\text{jumps}:~~~1\rightarrow 2,~~2\rightarrow 3,~~3\rightarrow 4,~~4\rightarrow 2.
$$   
In Fig.~\ref{fig1}(b), all these CW jumps, are shown.
After defining CW rotations, we  assume that the probability for any CW jump is given by a 
response $S(\vec{r},t)$ from a chemotactic memory, and other jumps are determined randomly so that:
\begin{equation}
P_{ij}= \left\{
\begin{array}{lr}
S(\vec{r},t)     \text{,}    ~~~~~~~~~~ \text{CW}~ \text{jumps}    \\
\frac{1-S(\vec{r},t) }{2}       \text{,}  ~~~~~~~~~\text{otherwise}.     \\
\end{array} \right.
\label{e-6}
\end{equation}
Here $\vec{r}$, is the position of the walker at time $t$.
Similar to the 
chemotaxis signaling network of E. coli, we assume that the signal $S$, is connected to the source 
$c$ (the local concentrations of food) through an intermediate dimensionless memory function $m$ as:
$S(\vec{r},t)=\xi/(1+\exp[m(\vec{r},t)-v_0c(\vec{r})])$ \cite{vladimirov1,vladimirov2}. The dynamics of memory function is given by: 
$\dot{m}(\vec{r},t)=(\tau^{H}/\tau_{ch})(S(\vec{r},t)-\xi/3)$,
where the dimensionless time scale of the adaptation is given by $\tau_{ch}/\tau^{H}$ 
and $v_{0}$ is a constant that has the dimension of volume.  For a uniform profile of the concentration, this system reaches a steady state with $S^{*}(\vec{r},t)=\xi/3$. Here
$\xi$ is a parameter in the interval $]0,1]$ and it shows how the internal jumps of the walker 
is anisotropic in the absence of any food gradient. Throughout this paper we will choose $\xi=0.95$.    
For a 
nonuniform concentration profile, there is no static steady state solution and 
the system evolves in time by continuously adjusting  its relative position and 
orientation with respect to the concentration profile.  

The statistical properties of this swimmer will be studied in details in the next section.
\begin{figure}
\includegraphics[width=0.9\columnwidth]{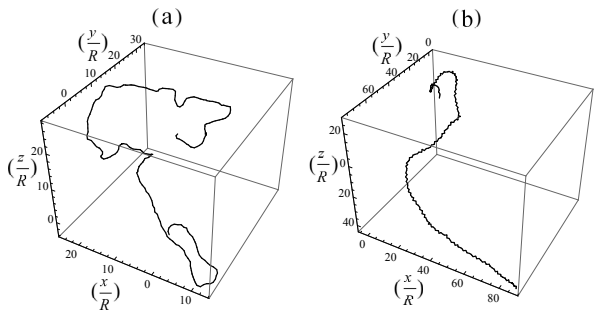}
\caption{Two different trajectories for the walker are shown. Part (a) corresponds to a 
walker moving in a unifrom concentration and (b) a linear concentration. 
Other numerical parameters are: $ a=0.2R$, $L=6.1R$, 
$\varepsilon=0.6R$, $\varphi=\pi /6 $, $\tau^H=0.02$, $ \tau^{ch}=100\tau^H$.}
\label{fig2}
\end{figure}

\begin{figure}
\includegraphics[width=0.95\columnwidth]{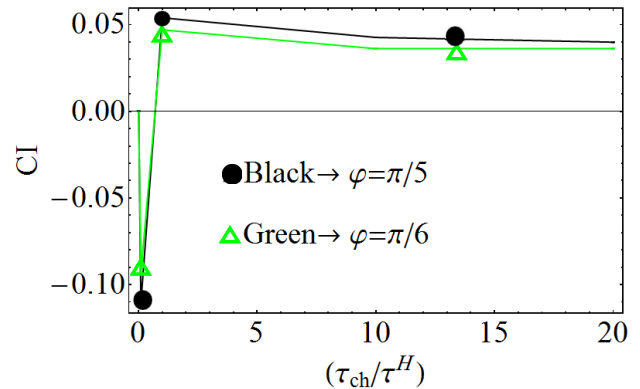}
\caption{Chemotactic index in terms of the memory time is investigated for two walkers with different 
geometrical values. The positive overcome of the chemotactic mechanism is not sensitive to 
the geometrical parameter.}
\label{fig3}
\end{figure}

\begin{figure*}
\includegraphics[width=0.99\linewidth]{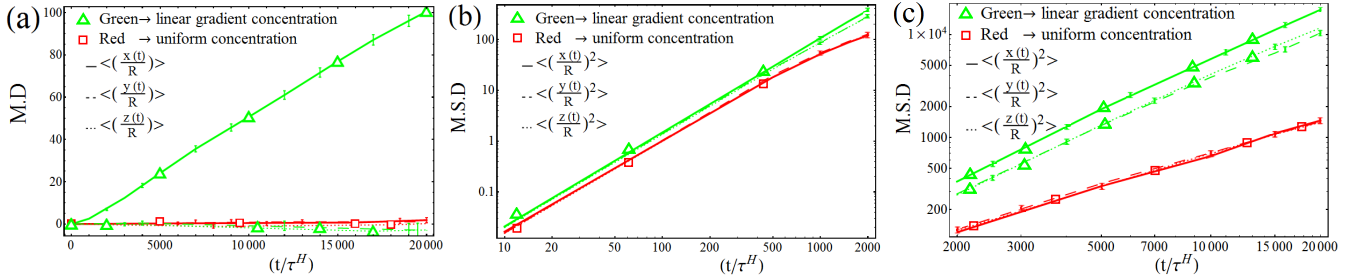}
\caption{Statistical properties of walkers moving in either a uniform concentration or a concentration with a linear gradient. 
Part (a) shows average position of the walker as a function of time. As one can see 
a concentration with gradient in the $x$ direction will result an average swimming to the positions with 
higher concentration of food. (b) shows  the short time behavior of mean square displacement 
in terms of time and (c) shows the corresponding behavior at large time scales. The chemotaxis memory shows positive results only at large time scales. For a walker moving in uniform concentration, 
the long time and short time behaviors are separated by a nonlinear crossover. 
This reflects the non symmetric nature of our passive walker. 
The numerical parameters are as in Fig.~\ref{fig2}.
}
\label{fig4}
\end{figure*}

\section{Results}
Fig.~\ref{fig2}(a) shows a typical trajectory of the walker in a uniform 
concentration profile of food molecules. It represents the trajectory of a random walker. 
To study the effect of 
a nonuniform concentration we choose a linear gradients in $x$ direction. Throughout this paper, 
and for nonuniform concentration, we choose a linear gradient with 
$v_0c(\vec{r})=100 x/R$.  A typical trajectory for the walker 
moving in this concentration field is shown in Fig.~\ref{fig2}(b). 
As one can see, the subsequent tumbles bias the  
trajectory toward the place with larger concentration of food molecules.   
In the literature of chemotaxis, $CI$ the chemotactic index is an important quantity that  
shows how accurate is a direction sensing mechanism. It is defined as the ratio of the walking displacement along the concentration gradient to the total length of the walking trajectory.  Depending on the dynamical variables of the system, $CI$ belongs to the interval $[-1,1]$. 
In Fig.~\ref{fig3}, $CI$ is plotted in terms of the dimensionless 
memory time $\tau_{ch}/\tau^H$. Here $\tau_{ch}$ is a parameter that comes from the 
chemical dynamics and  $\tau^{H}$ is a geometrical parameter.
 For large memory times ($\tau_{ch}\geq\tau^H$), the chemotactic index 
is positive. This is a signature saying that the chemotactic mechanism has a positive overcome and the walker can successfully  reach the target, the place of more food. 
As the time scale for a single jump is given by $\tau^{H}$, this proves that $\tau_{ch}$ plays the 
role of  a 
memory time. The memory time should be greater than the individual jumping time and this is 
the only condition required to have a successful gradient sensing  walker.    
The $CI$ is calculated for two different values of the apex angle. It is seen that the overcome of the searching mechanism is not so sensitive to  this angle. 
Now we can  study the statistical properties of the system for $\tau_{ch}\geq\tau^H$.
To have a better understanding of the role of fluctuations, we 
repeat the simulations for an ensemble of walkers and have studied the average 
statistical properties of the system. 
Mean displacement (MS), mean square displacements (MSD) and correlation functions are the statistical 
variables that we consider.    
To quantify the results of MSD, we define the diffusion exponents as: 
$$
\langle x^2\rangle\sim t^{\nu_\parallel},~~~
\langle y^2\rangle=\langle z^2\rangle\sim t^{\nu_\perp},
$$
where $\nu_{\parallel}$ and $\nu_{\perp}$ are the diffusion exponents along the gradient  and 
along a  direction perpendicular to the gradient, respectively. 
For a symmetric and normal random walker we have: $\nu_\parallel=\nu_\perp=1$. 
MD for a random walker moving in uniform and nonuniform concentration profiles 
are presented in Fig.~\ref{fig4}(a). As we expect, for nonuniform concentration the characteristics 
of a random walker is recovered, but for a  nonuniform concentration with a gradient in $x$ direction, 
the walker is biased  toward the positive $x$ direction. For nonuniform concentration,  
MD in the perpendicular directions ($\langle y\rangle$, $\langle z\rangle$)are zero but it is not 
zero  in the direction of gradient 
($\langle x\rangle$).

\begin{figure}
\includegraphics[width=0.9\columnwidth]{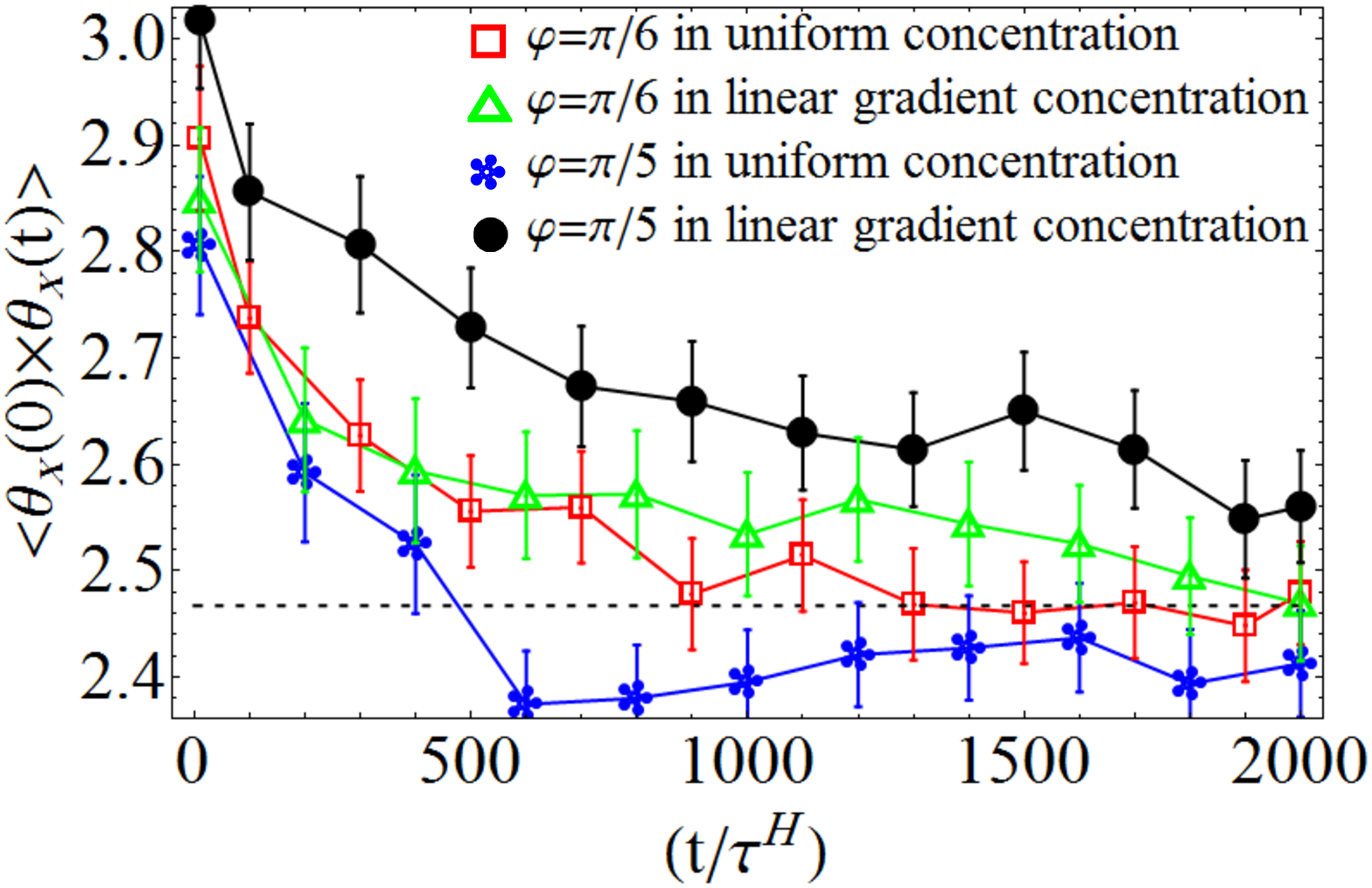}
\caption{Orientational correlation function is plotted as a function of time. Here $\theta_x(t)$, 
is the angle that the director vector of walker makes with the $x$ axis of the laboratory frame. 
Correlation time is sensitive on the geometrical variable (here $\varphi$) of the walker. 
For larger apex angle, the correlation function for a nonuniform gradient is larger 
than the corresponding value in uniform concentration. The dashed line shows the complete uncorrelated 
state with correlation $\pi^2/4$. The numerical parameters are as in Fig.~\ref{fig2}.}
\label{fig5}
\end{figure}

MSD in terms of time, in logarithmic scale, shows a nonlinear crossover from a short time to long time behavior. 
Figures~\ref{fig4}(b) and (c) show the short and long time 
MSD results   for this walker. This crossover is a result of the hydrodynamical anisotropy 
of the system. Please note that our system, spherical body 
with a  connected tail, is  anisotropic.  A similar crossover is recently observed for a diffusing 
object with boomerang geometry \cite{boomerang}. Short time behavior for a walker moving in a 
linear concentration corresponds to $\nu_\parallel=1.8$ and $\nu_\perp=1.7$.  
Long time behavior of the walker moving in a uniform concentration shows $\nu_\parallel=\nu_\perp=1.0$, 
that characterizes  a normal random walker (Fig.~\ref{fig4} (c)). For a walker moving in a nonuniform concentration, the exponents are given by $\nu_\parallel=1.6$ and $\nu_\perp=1.5$. 
The walker performs a super diffusion in all directions with an asymmetry 
in the direction parallel to the gradient concentration. To have more insights about the role of the geometry, we have studied the orientational 
correlation function $\langle\theta_{x}(0)\theta_{x}(t)\rangle$ where  $\theta_x(t)$ 
represents the angle that 
the director of the walker makes with $x$ axis (parallel to the concentration gradients). 
Fig.~\ref{fig5}, shows the features of the  correlation function for walkers moving in uniform and nonuniform concentrations. Correlation time, the decay time for the correlation, is sensitive to the 
apex angle. For larger apex angles,  the correlation time is also larger. This graph shows that the 
crossover time is essentially the time that orientational correlation washes out.

In conclusion, we have considered the statistical properties of a model hydrodynamical walker moving 
in a gradient field of food. 
Nontrivial coupling of geometrical parameters with dynamical parameters, reveals interesting 
statistical properties of the walker.   
 The memory time that is introduced within chemotactic mechanism, makes a 
superdiffusion walker. A crossover  from a short time to long time behavior of MSD is observed and the crossover time is the orientational correlation time.  The  hydrodynamic interaction 
between different walkers, is shown to have interesting features \cite{coherentcoupling}. 
Along the extension of this work, we  are  considering the role of hydrodynamic couplings in the physics 
of chemotaxis . 

\acknowledgements
A. N.  acknowledges  the Abdus Salam
international centre for theoretical physics for hospitality during the final  stage of this work.


\begin{thebibliography}{99}
\bibitem{cwrotation}
Please note that this definition for the CW rotations is not unique.   Other choices can work well.
\end{thebibliography}
\end{document}